\documentclass[10pt,%
aps,pra,
superscriptaddress]{revtex4-2}

\usepackage{amsthm}
\usepackage{amsmath}
\usepackage{mathrsfs}
\usepackage{amssymb} 
\usepackage{bbm,dsfont} 
\usepackage{multirow} 
\usepackage{units}
\usepackage{stmaryrd}   
\usepackage{verbatim}
\usepackage{enumerate} 
\usepackage{tikz}
\usetikzlibrary{calc}
\usepackage{tabularx}

\usepackage{dcolumn}

\usepackage{xcolor}
\usepackage{hyperref} %
\usepackage%
{hyperref} %
\hypersetup{
    colorlinks,
    linkcolor={blue!80!black},%
    citecolor={green!50!black},
    urlcolor={blue!80!black}
}

\usepackage{graphicx}

\newcommand{\bieee}{\begin{IEEEeqnarray}{rCl}}
\newcommand{\eieee}{\end{IEEEeqnarray}}
\newcommand{\prob}[1]{\Pr\left(#1\right)}
\newcommand{\given}{\mid}
\newcommand{\cprob}[2]{\Pr\left(#1\given #2\right)}

\renewcommand{\mathbbm}[1]{\text{\usefont{U}{bbm}{m}{n}#1}} %
\newcommand{\eps}{\varepsilon}

\newcommand{\norm}[1]{\left\lVert#1\right\rVert}
\newcommand{\trace}{\mathrm{Tr}}

\newcommand{\identity}{\mathbbm{1}}
\newcommand{\kb}[1]{ | #1 \rangle\langle #1 | } %

\newcommand{\ie}{\emph{i.e.} }

\newcommand{\etal}{\emph{et al.} }

\newcommand{\tR}{\widetilde{R}}

\newcommand{\hm}{\hat{m}}

\newcommand{\hP}{\hat{P}}
\newcommand{\hM}{\hat{M}}

\newcommand{\Aset}{\mathcal{A}}

\newcommand{\Dset}{\mathcal{D}}
\newcommand{\Fset}{\mathcal{F}}

\newcommand{\Hset}{\mathcal{H}}

\newcommand{\Mset}{\mathcal{M}}
\newcommand{\Pset}{\mathcal{P}}

\newcommand{\Sset}{\mathcal{S}}

\newcommand{\Xset}{\mathcal{X}}
\newcommand{\Yset}{\mathcal{Y}}
\newcommand{\Zset}{\mathcal{Z}}
\newcommand{\Eset}{\mathcal{E}}

\theoremstyle{remark}	\newtheorem{theorem}{Theorem}
\theoremstyle{remark}	\newtheorem{lemma}[theorem]{Lemma}
\theoremstyle{remark}	\newtheorem{corollary}[theorem]{Corollary}
\theoremstyle{remark}	
\theoremstyle{remark} \newtheorem{definition}{Definition}
\theoremstyle{remark} \newtheorem{remark}{Remark}
\theoremstyle{remark} \newtheorem{example}{Example}

\newcommand{\pSpace}{\mathcal{P}}														%

\newcommand{\tset}{\Aset^{\delta}}													%
\newcommand{\Tset}{\mathcal{T}}												%

\newcommand{\channel}{\mathcal{N}}

\newcommand{\inR}{\mathsf{R}}

\newcommand{\opR}{\mathbb{R}}
\newcommand{\opC}{\mathbb{C}}

\begin{document}
\title{Classical State Masking over a Quantum Channel} 

\author{Uzi Pereg}
\email{uzi.pereg@tum.de}
\affiliation{Institute for Communications Engineering, Technical University of Munich}%
\affiliation{
 Munich Center for Quantum Science and Technology (MCQST)
}%
\author{Christian Deppe}%
 \email{christian.deppe@tum.de}
\affiliation{Institute for Communications Engineering, Technical University of Munich}%

\author{Holger Boche}
 \email{boche@tum.de}
\affiliation{
 Theoretical Information Technology, Technical University of Munich}%
\affiliation{
 Munich Center for Quantum Science and Technology (MCQST)
}%
\affiliation{
Cyber Security in the Age of Large-Scale Adversaries Exzellenzcluster (CASA)
}%

\begin{abstract} 
Transmission of classical information over a quantum state-dependent channel is considered, when the encoder can measure channel side information (CSI) and is required to mask information on the quantum channel state %
from the decoder. In this quantum setting, it is essential to conceal the CSI measurement as well.
A regularized formula is derived for the  masking equivocation region, and a full characterization is established for a class of measurement channels.
\end{abstract}

\maketitle

\section{Introduction}
Security and privacy are critical aspects in modern communication systems 
\cite{BHCPDA:13p,LRBW:17p,PiquerasJoverMarojevic:19p,WYDP:19p}. 
In Wyner's wiretap  setting \cite{Wyner:75p}, the sender transmits a sequence $X^n$ over a memoryless broadcast channel $p_{Y,Z|X}$, such that the output sequence $Y^n$ is decoded by the legitimate receiver, while $Z^n$ is received by a malicious eavesdropper. Confidentiality requires that the eavesdropper cannot obtain information on the transmitted message from the sequence $Z^n$.
On the other hand, 
Merhav and Shamai \cite{MerhavShamai:07p} introduced a communication system with the privacy requirement of masking.

In the classical masking setting, the sender transmits a sequence $X^n$ over a memoryless state-dependent channel $p_{Y|X,S}$, where the state sequence $S^n$ has a fixed memoryless distribution and is not affected by the transmission.  The transmitter of $X^n$ is informed of $S^n$ and is required to send information to the receiver while limiting the amount of information that the receiver can learn about $S^n$.  
Intuitively, as the transmitter uses the side information in order to increase the transmission rate, more information on the channel state may be revealed. Hence, there is a tradeoff between high transmission rate and low leakage of information \cite{MerhavShamai:07p}.
The masking setting can also be viewed as communication with an untrusted party, where Alice wishes to send Bob a limited amount of information, while keeping the information source hidden \cite{Naor:91p,NaorPinkas:01p,JensenLuYiu:09b,SemalMarkantonakisAkram:18c,BFGMMS:21p}. 
It is expected that %
protocols that can solve communication tasks even under untrusted hardware platforms or untrusted software implementations will play an important role in the development of future communication systems \cite{FettweisBoche:21m,FettweisBoche:22p}.
Related settings and extensions are also considered in \cite{LeTreustBloch:16c,LeTreustBloch:20p,KoyluogluSoundararajanVishwanath:16p,KoyluogluSoundararajanVishwanath:11c,DikshteinSomekhBaruchShamai:19c,AsoodehDizaAlajajiLinder:16p,TutunchuogluOzelYenerUlukus:14c,Courtade:12c}.

Quantum information technology is rapidly evolving in both practice and theory 
\cite{NCLMSYH:20p%
}.
Communication through quantum channels can be separated into different categories. In particular, in  quantum information theory and Shannon theory, the following models of communication over quantum channels are considered in the literature:
\begin{enumerate}[A)]
\item
Transmission of classical information without assistance.
\item
Subspace transmission without assistance.
\item
Communication with entanglement assistance.
\end{enumerate}

 For classical communication without assistance, \mbox{model A,} the Holevo-Schumacher-Westmo- reland (HSW) Theorem provides a regularized (``multi-letter")  formula for the capacity of a quantum channel %
\cite{Holevo:98p,SchumacherWestmoreland:97p}. %
Although calculation of such a formula is intractable in general, it provides computable lower bounds, and there are special cases where the capacity can be computed exactly \cite{Shor:02p,Holevo:12b}. 
The reason for this difficulty is that the Holevo information is not necessarily additive \cite{Holevo:12b}. %
A similar difficulty occurs in \mbox{model B}, treating the transmission of quantum information \cite{Devetak:05p}.

Model C above is a scenario where
Alice and Bob have access to entanglement resources that are shared a priori, before communication takes place.  %
While entanglement can be used to produce shared randomness, it is a much more powerful aid \cite{BFSDBFJ:20b}. %
E.g., using super-dense coding, %
entanglement assistance doubles the transmission rate of classical messages over a noiseless qubit channel. 
 The entanglement-assisted capacity of %
a noisy quantum channel was fully characterized by Bennett \etal 
\cite{%
BennettShorSmolin:02p} in terms of the quantum mutual information. 
Entanglement resources are thus instrumental for the %
analysis of quantum communication systems,  providing %
a computable upper bound for unassisted communication as well.

Boche, Cai, and N\"{o}tzel \cite{BocheCaiNotzel:16p} addressed classical-quantum channels  with channel side information (CSI) at the encoder. %
The %
capacity %
was determined given causal CSI, and a regularized formula was provided given %
non-causal CSI \cite{BocheCaiNotzel:16p}.
The first author \cite{Pereg:20c1,Pereg:21p} extended the results to a quantum-input quantum-output channel with random parameters, and further considered communication over quantum channels with parameter estimation at the receiver, given either strictly-causal, causal, or non-causal CSI at the encoder, and without CSI as well.
Warsi and Coon \cite{WarsiCoon:17p} used an information-spectrum approach to derive  multi-letter bounds for a similar setting with %
rate-limited CSI. The entanglement-assisted capacity of a quantum channel with non-causal CSI was determined by    Dupuis in \cite{Dupuis:08a,Dupuis:09c} (see also \cite{AnshuJainWarsi:19p}), and with causal CSI  in \cite{Pereg:19c3,Pereg:19a}. 
 Luo and Devetak \cite{LuoDevetak:09p} considered channel simulation with source side information (SSI) at the 
decoder, and also solved the quantum generalization of the Wyner-Ziv problem \cite{WynerZiv:76p}.  Quantum data  compression with SSI  is also studied in \cite{DevetakWinter:03p,YardDevetak:09p,HsiehWatanabe:16p,DattaHircheWinter:19c,DattaHircheWinter:18a,
CHDH:19c,%
KhanianWinter:20p%
}.

Considering secure communication over the quantum wiretap channel,
 Devetak \cite{Devetak:05p} and Cai \etal \cite{CaiWinterYeung:04p}  established a regularized characterization of the secrecy capacity   without assistance. 
Related models appear in 
\cite{HsiehLuoBrun:08p,LiWinterZouGuo:09,Wilde:11p,Watanabe:12p,ElkoussStrelchuk:15p,AnshuHayashiWarsi:18c} as well.
Boche \etal \cite{BocheCaiNotzelDeppe:19p,BochCaiDeppeNotzel:17p} studied the quantum wiretap channel with an active jammer.
The capacity-equivocation region, %
characterizing the tradeoff between secret key consumption and private classical communication, was established in \cite{HsiehLuoBrun:08p,Wilde:11p}. %
The quantum 
Gel'fand-Pinsker wiretap channel is considered in \cite{AnshuHayashiWarsi:18c} and
other related scenarios can be found in \cite{KonigRennerBariskaMaurer:07p,GHKLLSTW:14p,LupoWildeLloyd:16p}.
Furthermore, network settings with confidential messages were recently considered in  \cite{%
SalekHsiehFonollosa:19c,%
AghaeeAkhbari:19c,BochJanssenSaeedianaeeni:20p}, respectively. %

\begin{figure}[tb]
\includegraphics[scale=0.7,trim={0 9.5cm 0 8.5cm},clip]{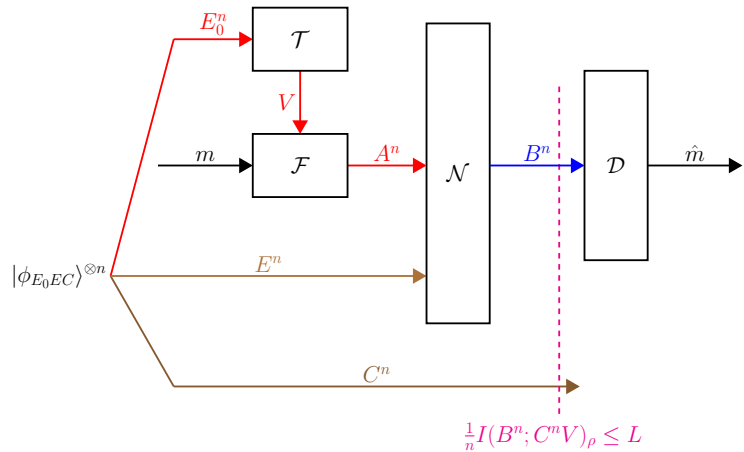} %
\\
\caption{Coding for a quantum state-dependent channel $\channel_{EA\rightarrow B}$ given side information at the encoder and masking from the decoder. The quantum systems of Alice and Bob are marked in red and blue, respectively. The channel state systems $E^n$ and $C^n$ are marked in brown.
Alice wishes to send a classical message $m$ to Bob.
She has access to side-information systems $E_0^n$, which are entangled with the channel state systems $E^n$. Alice performs  a measurement $\Tset$, and obtains a measurement outcome $V$. 
Then, she applies an encoding map $\Fset: (m,V)\to \rho_{A^n}$, and transmits the system $A^n$ over the channel.
 Bob receives the channel output system $B^n$, and applies the decoding measurement $\Dset: \rho_{B^n} \to \hm$ to obtain an estimate $\hm$ for Alice's message, as a measurement outcome. 
A leakage rate $L$ is achieved if 
$
 \frac{1}{n} I(B^n;C^n V)_\rho \leq L %
$.
}
\label{fig:mskCl1}
\end{figure}

In quantum channel state masking,
analogously to the classical model  \cite{MerhavShamai:07p}, the channel state system $C$ %
store %
undesired quantum information which leaks
to the receiver. This can model a leakage of private network information to the end-user. %
Alternatively, $C^n$ may represent a separate transmission to another receiver (Charlie), in a product state, out of our control, and which is not intended to our receiver (Bob),
and is therefore to be concealed from him. Thus, the goal of the transmitter (Alice)  is to mask this undesired information as
much as possible on the one hand, and to transmit reliable information on the other. 
Masking can also be viewed as a building block for  cryptographic problems of oblivious transfer of information and  secure computation by untrusting parties. 
In a recent paper by the authors \cite{PeregDeppeBoche:21p},  we considered a quantum state-dependent channel, %
when the encoder has CSI and is required to mask information on the quantum channel state from the decoder.
We have established a full characterization  for the entanglement-assisted  masking region with maximally correlated channel state systems, and a regularized formula  for the quantum masking region without assistance. 
That is, we addressed model B and model C for quantum channel state masking in \cite{PeregDeppeBoche:21p}.

In this paper, we consider model A of a quantum state-dependent channel $\channel_{EA\rightarrow B}$, when the encoder has CSI and is required to mask information on the quantum channel state from the decoder. We  derive a regularized formula for the classical  masking  region and establish full characterization for a class of measurement channels. Here, however,  the communication task is to send classical information, while there are no entanglement resources available to Alice and Bob. 
Specifically, the channel state systems are in an entangled state $|\phi_{E_0 EC}\rangle^{\otimes n}$.  Alice wishes to send a classical message  $m$. To this end, she measures the CSI systems $E_0^n$ and obtains an outcome $V$. Based on the measurement outcome, Alice encodes the quantum state of the channel input systems $A^n$ in  such a manner that limits the leakage-rate of Bob's information on $C^n$ from $B^n$,
while the systems  $E_0^n$ and $C^n$ are entangled with the channel state systems $E^n$ (see Figure~\ref{fig:mskCl1}). %

The quantum model involves three channel state systems, $E^n$, $E_0^n$, and $C^n$, as opposed to the classical case \cite{MerhavShamai:07p} of a single random parameter.
The system $E_0^n$ can be thought of as part of the environment of both our transmitter and the source of $C^n$, possibly entangled if they had previous interaction, while $E^n$ belongs to the channel's environment. 
The interpretation given in \cite{Dupuis:09c}, for the entanglement between $E_0^n$ and $E^n$, is that Alice shares entanglement with the channel itself. 
Another distinction from the classical case is that the measurement can cause a collapse of the wave function, hence correlations can be lost. Thereby, it is essential to conceal the CSI observation as well.
In the present model, the leakage requirement involves both the masked system $C^n$ and the measurement outcome $V$. Those subtleties do not exist in the classical problem.

Compared to our previous work \cite{PeregDeppeBoche:21p}, we now address a more fundamental problem in the following sense. In model A, we consider  a classical task, \ie the transmission of classical bits, that is performed using a quantum apparatus.
The techniques in the analysis are significantly different as well. The proof in \cite{PeregDeppeBoche:21p} is based on the decoupling approach \cite{HaydenHorodeckiWinterYard:08p}, using a code that decouples both Bob's environment and the channel state systems from the input reference. %
Here, the analysis is based on the quantum packing lemma \cite{HsiehDevetakWinter:08p}, using type-class projectors and the classical binning technique, along with non-trivial arguments %
to establish the leakage requirement.

\section{Definitions and Related Work}
\subsection{Notation, States, and Information Measures}
 We use the following notation conventions. %
Calligraphic letters $\Xset,\Yset,\Zset,...$ are used for finite sets.
Lowercase letters $x,y,z,\ldots$  represent constants and values of classical random variables, and uppercase letters $X,Y,Z,\ldots$ represent classical random variables.  
 The distribution of a  random variable $X$ is specified by a probability mass function (pmf) 
	$p_X(x)$ over a finite set $\Xset$. %
 We use $x^j=(x_1,x_{2},\ldots,x_j)$ to denote  a sequence of letters from $\Xset$. %
 A random sequence $X^n$ and its distribution $p_{X^n}(x^n)$ are defined accordingly. 

The state of a quantum system $A$ is given by a density operator $\rho$ on the Hilbert space $\Hset_A$.
The state is said to be pure if $\rho=\kb{\psi}$, for some vector $|\psi\rangle\in\Hset_A$, where
$\langle \psi |$ %
is the Hermitian conjugate of $|\psi\rangle$. 
A measurement of a quantum system is any set of operators $\{\Lambda_j \}$ that forms a positive operator-valued measure (POVM), \ie
the operators are positive semi-definite and %
$\sum_j \Lambda_j=\identity$, where $\identity$ is the identity operator. %
According to the Born rule, if the system is in state $\rho$, then the probability of the measurement outcome $j$ is given by $p_A(j)=\trace(\Lambda_j \rho)$.

Define the quantum entropy of the density operator $\rho$ as
\begin{align}
H(\rho) &\triangleq -\trace[ \rho\log(\rho) ] \,.
\end{align}
We may also consider the state of a pair of systems $A$ and $B$ on the tensor product $\Hset_A\otimes \Hset_B$ of the corresponding Hilbert spaces.
Given a bipartite state $\rho_{AB}$, %
define the quantum mutual information as
\begin{align}
I(A;B)_\rho=H(\rho_A)+H(\rho_B)-H(\rho_{AB}) . %
\end{align} 
Furthermore, the conditional quantum entropy and mutual information are defined by
$H(A|B)_{\rho}=H(\rho_{AB})-H(\rho_B)$ and
$I(A;B|C)_{\rho}=H(A|C)_\rho+H(B|C)_\rho-H(A,B|C)_\rho$, respectively.

A pure bipartite state %
is called \emph{entangled} if it cannot be expressed as the tensor product %
of two states %
in $\Hset_A$ and $\Hset_B$. %
The maximally entangled state %
between two systems %
of dimension $D$ %
is defined by
$%
| \Phi_{AB} \rangle = \frac{1}{\sqrt{D}} \sum_{j=0}^{D-1} |j\rangle_A\otimes |j\rangle_B %
$, where $\{ |j\rangle_A \}_{j=0}^{D-1}$ and $\{ |j\rangle_B \}_{j=0}^{D-1}$  %
are respective orthonormal bases. %
Note that $I(A;B)_{\kb{\Phi}}=2\cdot \log(D)$ and $I(A\rangle B)_{\kb{\Phi}}= \log(D)$.

\subsection{Quantum Channel}
\label{subsec:Qchannel}
A quantum channel maps a quantum state at the sender system to a quantum state at the receiver system. 
Here, we consider a channel with two inputs, where one of the inputs, which is referred to as the channel state, is not controlled by the encoder.
Formally, a quantum state-dependent channel $(\channel_{E A\rightarrow B} ,|\phi_{E E_0 C}\rangle)$ is defined by a   linear, completely positive, trace preserving map 
$%
\channel_{E A\rightarrow B}  %
$ %
and a quantum state $|\phi_{E E_0 C}\rangle$.
This model can be interpreted as if the channel is entangled with the systems $E$, $E_0$, and $C$.

We assume that both the channel state systems and the quantum channel have a product form. That is, the joint state of 
the systems $E^n=(E_1,\ldots,E_n)$, $E_0^n=(E_{0,1},\ldots,E_{0,n})$ and $C^n=(C_1,\ldots,C_n)$ is $%
|\phi_{E E_0 C}\rangle^{\otimes n}$, and if the systems $A^n=(A_1',\ldots,A_n')$ are sent through $n$ channel uses, then the input state $\rho_{ E^n A^n}$ undergoes the tensor product mapping
$%
\channel_{E^n A^n\rightarrow B^n}\equiv  \channel_{E A\rightarrow B}^{\otimes n} %
$. %
Given CSI, the transmitter can measure the systems $E_0^n$, which are entangled with the channel state systems $E^n$.
We will further consider a secrecy requirement that limits the information that the receiver can obtain on $C^n$.
The sender and the receiver are often referred to as Alice and Bob. 

\begin{remark}
\label{rem:mixed}
Our results apply to the case where $E$, $E_0$, and $C$ are in a mixed state as well. Specifically, given a mixed state $\varphi_{E E_0 C}$, there exists a purification $| \phi_{GE E_0 C} \rangle$, such that the reduced density operator for this purification is $\varphi_{E E_0 C}$. Hence, we can redefine the channel as follows. First, replace the channel state system $E$   by $\tilde{E}=(G,E)$, and then consider the quantum state-dependent channel
$\widetilde{\channel}_{\tilde{E} A\rightarrow B}$, where
\begin{align}
\widetilde{\channel}_{GE  A\rightarrow B}(\rho_{GEA})=\channel_{EA\rightarrow B}(\trace_G(\rho_{GEA})) .
\label{eq:channelTm}
\end{align}
\end{remark}

We will also consider the quantum-classical special case. 
\begin{definition}
A  measurement channel (or, q-c channel) $\Mset_{A\rightarrow Y}$ has the following form,
\begin{align}
\Mset_{A\rightarrow Y}(\rho_{A})=  \sum_{y\in\Yset} \trace( \Lambda_{y} \rho_A) \kb{y}
\end{align}
for some POVM $\{\Lambda_y \}$ and orthonormal vectors $\{ |y\rangle \}$.
In order to distinguish it from the general channel, %
we  denote the state-dependent measurement channel by %
$(\Mset_{E A\rightarrow Y} ,|\phi\rangle)$ . %
\end{definition}

One may also consider the special case where the channel state is fully described by a classical random parameter, \ie
$E\equiv E_0\equiv C\equiv S$ where
 $S\sim q(s)$ is a classical random variable. In this case, the channel can be viewed as a random selection from a collection of channels $\{ \channel^{(s)}_{A\to B} \}_{s\in\Sset}$.
This family of quantum state-dependent channels is of particular interest as it captures the notion of \emph{channel uncertainty}.
For the so-called `random-parameter quantum channel',  %
the availability of CSI at the encoder simply means that Alice knows the value of $S$.
We give  simple examples below.

\begin{example}
\label{example:depolC0}
The random-parameter depolarizing channel is defined as follows \cite[Example 3]{Pereg:21p}. Let $\channel_{SA\to B}$ be a quantum state-dependent channel that depends on a classical random parameter $S\in \{0,1,2,3\}$, hence $E_0\equiv E\equiv C\equiv S$. As pointed out above, such a random-parameter quantum channel can be viewed as a random selection from a set of channels, $\{\channel^{(s)}\}_{s=0,1,2,3.}$. Let
\begin{align}
\channel^{(0)}(\rho)&=\rho\\
\channel^{(1)}(\rho)&=\mathsf{X}\rho\mathsf{X}\\
\channel^{(2)}(\rho)&=\mathsf{Y}\rho\mathsf{Y}\\
\channel^{(3)}(\rho)&=\mathsf{Z}\rho\mathsf{Z}
\end{align}
with the following parameter distribution, 
\begin{align}
q(0)=1-\frac{3\eps}{4} \,,\; q(1)=q(2)=q(3)=\frac{\eps}{4}
\end{align}
 where $\eps\in (0,1]$ is a given constant and $\mathsf{X}$, $\mathsf{Y}$, $\mathsf{Z}$ are the qubit Pauli operators. In other words, %
the parameter $S_i$ chooses a Pauli operator that is applied to the $i$th input system. 
We note that without CSI, the average channel is the same as the standard depolarizing channel, \ie
\begin{align}
\overline{\channel}_{A\rightarrow B}(\rho)&\equiv \sum_s q(s)\channel^{(s)}(\rho)
\nonumber\\
&=\left(1-\frac{3\eps}{4} \right)\rho+\frac{\eps}{4}\left( \mathsf{X}\rho\mathsf{X}+\mathsf{Y}\rho\mathsf{Y}+\mathsf{Z}\rho\mathsf{Z}  \right)
\nonumber\\
&=(1-\eps)\rho+\eps\pi
\end{align}
 where $\pi=\frac{\identity}{2}$ is the maximally mixed state (see \cite[Section 4.7.4]{Wilde:17b}). Without CSI, the capacity can be significantly lower than $1$.
In particular, for $\eps=1$, the capacity without CSI is zero and Alice cannot send any information to Bob.

Knowing the parameter $s$, Alice can revert the operation of the channel by applying the corresponding Pauil operator. That is, Alice applies  $\channel^{(s)}$  locally in her encoding operation, and then sends the input state $\channel^{(s)}(\rho)$ through the channel. Hence, Bob receives $\channel^{(s)}\left( \channel^{(s)}(\rho) \right)=\rho$. In this manner, we effectively have a noiseless channel. Furthermore, the channel output has no correlation with the channel state $S$. 
Thereby, Alice can send $1$ information bit per transmission without leakage.
\end{example}

\begin{example}
\label{example:erNC0}
Consider a random-parameter qubit channel %
that depends on a classical random parameter $S\sim\text{Bernoulli}(\eps)$, %
such that %
\begin{align}
\channel^{(0)}(\rho)&=\rho\\
\channel^{(1)}(\rho)&=|\psi\rangle\langle\psi|
\end{align}
where $|\psi\rangle$ is a given state  in the same qubit space. 
We will return to this example in the sequel and show that if Alice uses the CSI in order to increase the transmission rate, then there may be leakage of information on $S^n$ to Bob (see  Example~\ref{example:erNC}).
\end{example}

\subsection{Coding}
\label{subsec:Mcoding}
We define %
a privacy masking code to transmit classical information over a quantum channel. 
With non-causal CSI, Alice can measure the systems $E_0^n$, which are entangled with the channel state systems $C^n E^n$.
We refer to $E_0^n$ as the CSI systems.

\begin{definition} %
\label{def:Mcapacity}
A $(2^{nR},n)$ classical masking  code with  CSI at the encoder consists of the following:    
A message set $[1:2^{nR}]$,  assuming that $2^{nR}$ is integer,
an encoding POVM, $\Tset \equiv \{T^{v}_{E_0^n}\}$, on the CSI system $E_0^n$,
an encoding map $\Fset: (m,v)\mapsto \rho_{A^n}$, and a decoding POVM
$\Dset \equiv \{D^{\hm}_{B^n}\}  $.

The communication scheme is depicted in Figure~\ref{fig:mskCl1}.  
The sender Alice has the systems  $E_0^n$ and $A^n$, and the receiver Bob has the systems $B^n$. 
Alice chooses a classical message $m\in [1:2^{nR}]$ uniformly at random, and wishes to send it to Bob.
To this end, she measures the CSI systems $E_0^n$, which are entangled with the channel state systems, using the measurement set $\Tset$, and obtains a measurement outcome $v$. 
Then, Alice encodes the classical message $m$ using the measurement outcome, and prepares the input state $\rho^{m,v}_{A^n}=\Fset(m,v)$. The average post-measurement input state is
\begin{align}
\bar{\rho}^m_{C^n E^n V A^n}= \sum_v \trace_{E_0^n}\left( T_{E_0^n}^{v}\phi_{CEE_0}^{\otimes n} \right)\otimes \kb{v} \otimes \rho^{m,v}_{A^n}
\end{align}
where $V$ is a classical register that stores the CSI-measurement outcome.

Alice transmits the systems $A^n$ over %
$n$ channel uses of $\channel_{EA\rightarrow B}$. Hence, the average output state is
\begin{align}
\rho_{C^n V B^n }^m=\channel_{E^n A^n\rightarrow B^n} (\bar{\rho}^m_{C^n V E^n A^n}) .
\end{align}

 Bob receives the channel output and applies the decoding measurement $\Dset$ to the output systems $B^n$,
 such that the measurement outcome $\hm$ is an estimate of the original message $m$. The average probability of error is 
\begin{align}
P_{e}^{(n)}(\Tset,\Fset,\Dset)= 
1-\frac{1}{2^{nR}}\sum_{m=1}^{2^{nR}}\trace\left(  D^{m}_{B^n} \rho^m_{B^n} \right) 
\end{align}
where $\rho_{B^n}^m=\trace_{C^n V}(\rho_{C^n V B^n}^m)$.
The masking leakage rate of the code $(\Tset,\Fset,\Dset)$ is defined as
\begin{align}
\ell^{(n)}(\Tset,\Fset,\Dset)\triangleq  \frac{1}{n} I(C^n V;B^n )_{\rho} 
\label{eq:elln}
\end{align}
where the mutual information is computed with respect to the average states, corresponding to a uniformly distributed message and the random   outcome $V$ of the CSI measurement at the encoder.
A $(2^{nR},n,\eps,L)$ masking code satisfies 
$%
P_{e}^{(n)}(\Tset,\Fset,\Dset)\leq\eps $  %
and $\ell^{(n)}(\Tset,\Fset,\Dset)\leq L$.  %
A rate-leakage pair $(R,L)$, where $R,L\geq 0$, is called achievable  if for every $\eps,\delta>0$ and sufficiently large $n$, there exists a 
$(2^{nR},n,\eps,L+\delta)$ masking code. 

 The classical masking region $\opR_{\text{CL}}(\channel)$ of the quantum state-dependent channel $\channel_{EA\to B}$ is defined as the set of achievable pairs $(R,L)$ with CSI at the encoder. 
Alternatively, one may fix the leakage rate and consider the optimal transmission rate. The classical capacity-leakage function 
$\opC_{\text{Cl}}(\channel,L)$ is defined as the supremum of achievable rates $R$ for a given leakage $L$. %
Note that $\opC_{\text{Cl}}(\channel,\infty)$ reduces to the standard definition of the classical capacity of a quantum channel, without a masking requirement. 

\end{definition}

\begin{remark}
\label{rem:nonTmaskL}
Observe that if $L\geq 2\log|\Hset_B|$, then the masking requirement trivially holds because $I(C^n V;B^n )_\rho\leq
2H(B^n)_\rho\leq 2n\log|\Hset_B|$. That is, if $L\geq 2\log|\Hset_B|$, then
$\opC_{\text{Cl}}(\channel,L)=\opC_{\text{Cl}}(\channel,\infty)$.
\end{remark}

\subsection{Related Work}
\label{subsec:Previous}
We briefly review known results for the case where there is no masking requirement.
First, consider a quantum channel which is not affected by a channel state, \ie $\channel_{EA\rightarrow B}(\rho_{EA})=
\Pset_{A\rightarrow B}(\trace_E(\rho_{EA}))$. 
Define %
\begin{align}
\chi(\Pset)\triangleq   \max_{p_X(x), |\phi_A^x\rangle } I(X;B)_\rho  %
\label{eq:HolevoChan}
\end{align}
 with $\rho_{XB}\equiv \sum_{x\in\Xset} p_X(x) \kb{x}\otimes \Pset( \kb{ \phi_A^x })$ and 
$|\Xset|\leq |\Hset_A|^2$.
The objective functional $I(X;B)_\rho$ is referred to as the Holevo information with respect to the ensemble 
$\{ p_X(x), \Eset( \kb{ \phi_A^x })  \}$ and the channel $\Pset_{A\rightarrow B}$, while the formula $\chi(\Pset)$ itself is sometimes referred to as the Holevo information of the channel  \cite{Wilde:17b}.
 Next, we cite the HSW Theorem, which provides a regularized capacity formula for a quantum channel that does not depend on a state.
\begin{theorem}  [see {\cite{Holevo:98p,SchumacherWestmoreland:97p}}]
\label{theo:ClNoSI}
The classical capacity of a quantum channel $\Pset_{A\rightarrow B}$ that does not depend on a channel state, without a masking requirement, is given by 
\begin{align}
\opC_\text{Cl}(\Pset,\infty)= \lim_{n\rightarrow \infty} \frac{1}{n} \chi \left( \Pset^{\otimes n} \right) \,.
\label{eq:ClNoMsk}
\end{align}
\end{theorem}
A single-letter characterization is an open problem for a general quantum channel.
Although calculation of a regularized formula is intractable in general, it provides a computable lower bound, and there are special cases where the capacity can be computed exactly
\cite{DevetakShor:05p}.

Next, we move to a quantum state-dependent channel %
with CSI at the encoder, in the special case where the state is a classical random parameter $S\sim q(s)$. As explained in Subsection~\ref{subsec:Qchannel}, the channel $\channel_{SA\rightarrow B}$ can be specified by a collection of channels $\{ \channel^{(s)}_{A\to B} \}$. Define %
\begin{align}
\mathsf{R}(\channel,\infty)\triangleq   \sup_{p_{X|S}(x|s), \varphi_A^x } [I(X;B)_\rho-I(X;S)]  %
\label{eq:inRnc}
\end{align}
where the supremum is over the conditional distributions $p_{X|S}$ and the collections of input states $\varphi_A^x$, such that given $S=s$, we have the state 
 $\rho_{XB|s}\equiv \sum_{x\in\Xset}  p_{X|S}(x|s) \kb{x}\otimes \channel_{A\to B}^{(s)}( \varphi_A^x )$.
\begin{theorem} [see {\cite{Pereg:20c1,Pereg:21p}}]
\label{theo:CeaNC}
The classical capacity of a random-parameter quantum channel $(\channel_{SA\rightarrow B},S\sim q(s))$, with CSI at the encoder and without a masking requirement, is given by
\begin{align}
\opC_{\text{Cl}}(\channel,\infty)= \lim_{n\rightarrow \infty} \frac{1}{n} \mathsf{R}\left(\channel^{\otimes n},\infty\right) .
\label{eq:QeaNoMskE}
\end{align}
\end{theorem}

\section{Main Results}
We state our results on the quantum state-dependent channel $\channel_{E A\rightarrow B}$ with masking.
We determine a regularized characterization of the masking region and 
capacity-leakage function, for the transmission of classical information.
Define
\begin{align}
&\mathcal{R}_{\text{Cl}}(\channel)=
\bigcup_{ \Lambda_{E_0}^s \,,\; p_{X|S} \,,\; \varphi_{A}^{x} }
\left\{ \begin{array}{rl}
  (R,L) \,:\;
	0\leq R &\leq  I(X;B)_\rho- I(X; S)  \\
  L   &\geq I(CS;XB)_\rho
	\end{array}
\right\}
\label{eq:calRCl}
\end{align}
where the union is over the POVMs $\{\Lambda_{E_0}^s\}$, the conditional distributions $p_{X|S}$, and the collections of input states $\varphi_A^{x}$, 
with 
\begin{align}
\rho_{ECSXA} &=\sum_{s\in\Sset}\sum_{x\in\Xset} p_{X|S}(x|s)  \trace_{E_0}(\Lambda_{E_0}^s \phi_{E_0 EC})\otimes \kb{s,x}\otimes \varphi_{A}^x
\intertext{and} 
\rho_{BCSX}&= 
\channel_{EA\to B}( \rho_{EACSX} ).
\end{align}

\begin{theorem}
\label{theo:MskCl}
$\,$
\begin{enumerate}[1)]
\item
The classical masking region  of a quantum state-dependent channel $\left(\channel_{EA\rightarrow B}, |\phi_{E E_0 C}\rangle \right)$ with CSI at the encoder is given by
\begin{align}
\mathbb{R}_{\text{Cl}}(\channel)&=\bigcup_{n=1}^{\infty}\frac{1}{n}\mathcal{R}_{\text{Cl}}(\channel^{\otimes n}) \,.
\end{align}

\item
For a measurement channel $\Mset_{EA\to Y}$ with a classical CSI system $E_0\equiv S$,
\begin{align}
\mathbb{R}_{\text{Cl}}(\Mset)&=\bigcup_{  p_{X|S} \,,\; \varphi_{A}^{x} }
\left\{ \begin{array}{rl}
  (R,L) \,:\;
	0\leq R &\leq  I(X;Y)- I(X; S)  \\
  L   &\geq I(CS;XY)_\rho
	\end{array}
\right\} \,.
\label{eq:calRClM}
\end{align}
\end{enumerate}
\end{theorem}
The proof of Theorem~\ref{theo:MskCl} is given in Appendix~\ref{app:MskCl}.

\begin{remark}
In Appendix~\ref{app:Alphabet}, we show that the union can be exhausted with cardinality $|\Xset|\leq (|\Hset_A|^2+1)|\Hset_E|$.
Hence, in principle, the region $\mathcal{R}_{\text{Cl}}(\channel)$ is computable. Nevertheless, for a general quantum channel, we have only obtained a regularized characterization. As mentioned in Section~\ref{subsec:Previous}, a single-letter capacity formula is an open problem, even for a point-to-point quantum channel without a channel state.
\end{remark}

Equivalently, we can characterize the  capacity-leakage function. 
 The following corollary is an immediate consequence of Theorem~\ref{theo:MskCl}.
\begin{corollary}
\label{coro:Msk}
$\,$
\begin{enumerate}[1)]
\item
The classical capacity-leakage function  of a quantum state-dependent channel $\left(\channel_{EA\rightarrow B}, |\phi_{E E_0 C}\rangle \right)$ with CSI at the encoder is given by
\begin{align}
\opC_{\text{Cl}}(\channel,L)&=\lim_{n\to\infty}\frac{1}{n}\sup_{ \Lambda_{E_0^n}^s \,,\; p_{X|S} \,,\; \varphi_{A^n}^{x} \,:\;
 I(C^n S;X B^n)_\rho\leq L} [I(X;B^n)_\rho-I(X;S)] \,.
\end{align}

\item
For a measurement channel $\Mset_{EA\to B}$ with a classical CSI system $E_0\equiv S$,
\begin{align}
\opC_{\text{Cl}}(\Mset,L)&=\sup_{  p_{X|S} \,,\; \varphi_{A}^{x} \,:\; I(CS;XY)_\rho\leq L}
[I(X;Y)- I(X; S) ] \,.
\end{align}
\end{enumerate}
\end{corollary}

To illustrate our results, we return to the channels in  Examples \ref{example:depolC0} and \ref{example:erNC0}. %
Example~\ref{example:depolC0} is a trivial example where there is no tradeoff between the transmission rate and the leakage. 
Specifically, Alice can transmit $1$ bit of information per transmission without leakage. 
Hence, the capacity-leakage region of the random-parameter depolarizing channel is given by
\begin{align}
\opC_{\text{Cl}}(\channel)=
\left\{
\begin{array}{lrl}
(R,L) \,:\; & R \leq& 1 \\
						& L \geq& 0
\end{array}
\right\} \,.
\end{align}
Now, we demonstrate the tradeoff for the channel in Example~\ref{example:erNC0}.

\begin{example}
\label{example:erNC}
Consider a qubit channel $\channel_{SA\to B}$ that depends on a classical random parameter $S\sim\text{Bernoulli}(\eps)$, hence $E_0\equiv E\equiv C\equiv S$. As pointed out in \ref{subsec:Qchannel}, such a random-parameter quantum channel can be viewed as a random selection from a set of channels, $\{\channel^{(s)}\}_{s=0,1.}$. Let
\begin{align}
\channel^{(0)}(\rho)&=\rho\\
\channel^{(1)}(\rho)&=|\psi\rangle\langle\psi|
\end{align}
where $|\psi\rangle$ is a given state  in the same qubit space, as in Example~\ref{example:erNC0}. 
Here, %
the parameter $S_i$ chooses whether the $i$th input system is projected onto $|\psi\rangle$.
This channel has also been considered in  the dual model of parameter estimation \cite[Example 4]{Pereg:21p}. 
Ignoring the CSI at the encoder, the average channel $\overline{\channel}_{A\to B}(\rho)=(1-\eps)\rho+\eps\kb{\psi}$ resembles the quantum erasure channel \cite{BennettDiVincenzoSmolin:97p} (see also \cite[Section 20.4.3]{Wilde:17b}), except that the ``erasure state"of an erasure channel is orthogonal to the qubit space, while $|\psi\rangle$ in the present example is in the same qubit space. 
Nonetheless, we note that if the decoder knows the locations where the state is projected, then this model is equivalent to the quantum erasure channel.
Without this knowledge at the decoder, it is less obvious.

By Theorem~\ref{theo:MskCl}, the following rate-leakage region is achievable for the random-parameter channel above,
\begin{align}
\opC_{\text{Cl}}(\channel)\supseteq 
\bigcup_{0\leq \alpha\leq \frac{1}{2}}
\left\{
\begin{array}{lrl}
(R,L) \,:\; & R \leq& (1-\eps)h(\alpha) \\
						& L \geq& h\left((1-\eps)\alpha\right)-(1-\eps)h(\alpha)
\end{array}
\right\} 
\end{align}
where $h(x)$  is the binary entropy function, \ie $h(x)=-(1-x)\log(1-x)-x\log(x)$ for $x\in (0,1)$, and $h(0)=h(1)=0$.
We can see the tradeoff between the communication rate and the leakage. 
Clearly, if the encoder constantly transmits  $|\psi\rangle$, then there is no leakage, as the output is $|\psi\rangle\otimes\cdots\otimes |\psi\rangle$. Yet, the rate is zero as well. 
Indeed, for $\alpha=0$, we achieve $(R,L)=(0,0)$.
On the other hand, taking $\alpha=\frac{1}{2}$, we obtain the maximal rate 
$R=1-\eps$, which is also the capacity of the quantum erasure channel. However, the leakage is $L=h\left(\frac{1}{2}(1-\eps)\right)-(1-\eps)$.

To show this, note that the bound on the rate on the RHS of (\ref{eq:calRCl}) can also be expressed as
\begin{align}
R&\leq H(X|S)-H(X|B)_\rho \nonumber\\
&= H(X|S)-H(XB)_\rho+H(B)_\rho
\,.
\end{align}
Given  CSI at the encoder, we can choose an auxiliary $X$ that depends on the channel parameter $S$.
Let the input ensemble be the basis $\{|\psi\rangle, |\psi_\perp\rangle\}$, where $|\psi_\perp\rangle$ is orthogonal with respect to $|\psi\rangle$. %
 The input distribution is chosen as follows. Let $V\sim \text{Bernoulli}(\alpha)$ be statistically independent of $S$.
If $S=0$, set $X=V$. Otherwise, if $S=1$, set $X=0$.
This results in the following quantum state,
\begin{align}
\rho_{SXB}&=(1-\eps)\kb{0}\otimes \left( (1-\alpha)\kb{0}\otimes \kb{\psi}+\alpha\kb{1}\otimes\kb{\psi_\perp}  \right)
\nonumber\\&
+\eps\kb{1}\otimes \kb{0}\otimes\kb{\psi}
\\
\rho_{XB}
&=[(1-\eps)(1-\alpha)+\eps]\kb{0}\otimes \kb{\psi}+(1-\eps)\alpha\kb{1}\otimes\kb{\psi_\perp}
\end{align}
Hence,
\begin{align}
 H(XB)_\rho&=H(B)_\rho=h\left((1-\eps)\alpha\right) \\
 H(XB|S)_\rho&=H(X|S)=(1-\eps)\cdot H(V)+\eps\cdot 0=(1-\eps)h(\alpha)
\intertext{and}
 I(S;XB)_\rho&=H(XB)_\rho-H(XB|S)_\rho=h\left((1-\eps)\alpha\right)-(1-\eps)h(\alpha)
\end{align}
\end{example}

\section{Summary and Concluding Remarks}
We consider  communication of classical information over a quantum state-dependent channel $\channel_{EA\to B}$, when the encoder can measure channel side information (CSI) and is required to mask information on the quantum channel state from the decoder. 
Specifically, the channel state systems are in an entangled state $|\phi_{E_0 EC}\rangle^{\otimes n}$  (see Figure~\ref{fig:mskCl1}).   Alice wishes to send a classical message  $m$. To this end, she measures the CSI systems $E_0^n$ and obtains an outcome $V$. Based on the measurement outcome, Alice encodes the quantum state of the channel input systems $A^n$ in  such a manner that limits the leakage-rate of Bob's information on $C^n$ from $B^n$. %

In quantum channel state masking,
analogously to the classical model \cite{MerhavShamai:07p}, the channel state system $C^n$   
store %
undesired quantum information which leaks
to the receiver. This can model a leakage of secret network information in the system to the end-user.
Alternatively, the  state system $C^n$ may represent another transmission to another receiver, Charlie, %
which is not intended to Bob, %
and is therefore to be concealed from him. Thus, Alice's goal  is to mask this undesired information as
much as possible on the one hand, and to transmit reliable  information on the other. 

In a recent paper by the authors \cite{PeregDeppeBoche:21p}, we have considered a quantum state-dependent channel $\channel_{EA\rightarrow B}$, when the encoder has CSI and is required to mask information on the quantum channel state from the decoder.
We have established a full characterization  for the entanglement-assisted masking  region with maximally correlated channel state systems, and a regularized formula  for the quantum masking region without assistance. 
Here, we have removed the entanglement assistance, and considered the transmission of \emph{classical} information over the quantum channel.

Masking can also be viewed as a building block for  cryptographic problems of oblivious transfer of information, such as  bit commitment or %
secure computation.
Suppose that Alice is a server that receives a query. She is required to use a quantum computer in order to compute a difficult task, while also using a private source $E_0^n C^n$.
To this end, Alice uses $E_0^n$ to encode $A^n$, including a reference number $m$ (metadata), which could possibly include the computation query as well.  
Next, she performs the computation map $\channel_{EA\to B}^{\otimes n}$ on the systems $E^n A^n$, which are entangled with the private source. The quantum output system $B^n$ is delivered to the agent Bob, who performs a measurement to view the metadata $m$, and then use $B^n$ as he wishes.
The masking requirement is to prevent Bob from recovering the server's private source.

\section*{Acknowledgment}
Uzi Pereg, Christian Deppe, and Holger Boche were supported by the Bundesministerium f\"ur Bildung und Forschung (BMBF) through Grants
16KISQ028 %
(Pereg, Deppe) and 16KIS0858 (Boche).
This work of H. Boche was supported in part by the German Federal
Ministry of Education and Research (BMBF) within the national initiative for
``Post Shannon Communication (NewCom)" under Grant 16KIS1003K and in
part by the German Research Foundation (DFG) within the Gottfried Wilhelm
Leibniz Prize under Grant BO 1734/20-1 and within Germany’s Excellence
Strategy EXC-2092 – 390781972 and EXC-2111 – 390814868. U. Pereg was also supported by the Israel CHE Fellowship for Quantum Science and Technology.

\appendix
\section{Cardinality Bound}
\label{app:Alphabet}
Consider the region $\mathcal{R}_{\text{Cl}}(\channel)$ as defined in (\ref{eq:calRCl}).
To bound the alphabet size, we use the Fenchel-Eggleston-Carath\'eodory lemma \cite{Eggleston:66p} and similar arguments as in
in \cite{Pereg:21p}. First, observe that since $\Lambda_{E_0}^s$ is a measurement on $E_0$, we can restrict the dimension of this measurement to $|\Hset_{E_0}|$, hence $|\Sset|\leq |\Hset_{E_0}|$.
Fix $p_S(s)=\trace(\Lambda_{E_0}^s\phi_{E_0})$, and 
consider the ensemble $\{  p_{X|S}(x|s) \,, \varphi_A^{x}  \}$. 
Every quantum state $\theta_A$ has a unique parametric representation $u(\theta_A)$ of dimension $|\Hset_A|^2-1$ (see \cite[Appendix B]{Pereg:21p}). 
Then, %
define a map $f_s:\Xset\rightarrow \mathbb{R}^{|\Hset_A|^2+1}$ by
\begin{align}
f_s(x)= \left(  u(\varphi_A^{x}) \,,\; -H(B|X=x)_\rho+H(S|X=x) \,,\; H(CS|B,X=x)_\rho   \right) \,.
\end{align}
The map $f_s$ can be extended to probability distributions as follows,
\begin{align}
F_s \,:\; p_{X|S}(\cdot|s)  \mapsto
\sum_{x\in\Xset} p_{X|S}(x|s) f_s(x)= \left(  u(\rho_A^s) \,,\; -H(B|X)_{\rho}+H(S|X) \,,\; H(CS|BX)_\rho   \right) %
\end{align}
for $s\in\Sset$, where $\rho_A^s=\sum_x p_{X|S}(x|s) \varphi_A^{x}$.
According to the Fenchel-Eggleston-Carath\'eodory lemma \cite{Eggleston:66p}, any point in the convex closure of a connected compact set within $\mathbb{R}^d$ belongs to the convex hull of $d$ points in the set. 
Since the map $F_s$ is linear, it maps  the set of distributions on $\Xset$ to a connected compact set in $\mathbb{R}^{|\Hset_A|^2+1}$. Thus, for every  $s$, 
there exists a conditional probability distribution $p_{\bar{X}|S}(\cdot|s)$ on a subset $\overline{\Xset}\subseteq \Xset$ of size $%
|\Hset_A|^2+1$, such that 
$%
F_s(p_{\bar{X}|S}(\cdot|s))=F_s(p_{X|S}(\cdot|s)) %
$. %
We deduce that the alphabet dimension can be restricted to $|\Xset|\leq (|\Hset_A|^2+1)|\Hset_{E_0}|$, while preserving $\rho_{SAEC}$ and
$\rho_{SBC}\equiv \channel_{EA\to B}(\rho_{SEAC})$; $I(X;B)_\rho-I(X;S)=H(B)_\rho-H(B|X)_{\rho}+H(S|X)-H(S)$; and
$I(CS;XB)_\rho=H(CS)_\rho-H(CS|BX)_\rho$.
\qed

\section{Information Theoretic Tools}
\label{sec:Itools}
To derive our results, we use the quantum version of the method of types properties and techniques. The basic definitions and lemmas are similar to those in \cite{Pereg:21p}. For convenience, we bring them here as well.

\subsection{Classical Types}
The type of a classical sequence $x^n$ is defined as the empirical distribution $\hP_{x^n}(a)=N(a|x^n)/n$ for $a\in\Xset$, where $N(a|x^n)$ is the number of occurrences of the symbol $a$ in the sequence $x^n$. Denote the set of all types over $\Xset$ is by 
$\pSpace_n(\Xset)$.
For a pair of sequences $x^n$ and $y^n$, we give similar definitions in terms of the joint type $\hP_{x^n,y^n}(a,b)=N(a,b|x^n,y^n)/n$ for $a\in\Xset$, $b\in\Yset$, where $N(a,b|x^n,y^n)$ is the number of occurrences of the symbol pair $(a,b)$ in the sequence 
$(x_i,y_i)_{i=1}^n$. Given a sequence $y^n\in \Yset^n$, we further define the conditional type $\hP_{x^n|y^n}(a|b)=N(a,b|x^n,y^n)/N(b|y^n)$.

Given a probability distribution $p_X\in\pSpace(\Xset)$, define  the $\delta$-typical set   as
\begin{align}
\tset(p_X)\equiv \bigg\{ x^n\in\Xset^n \,:\; %
\left| \hP_{x^n}(a) - p_X(a) \right|\leq\delta \quad\text{if $\, p_X(a)>0$}&  \nonumber\\ 
 \hP_{x^n}(a)=0 \quad\text{if $\, p_X(a)=0$} &, \;\text{$\forall$ $a\in\Xset$} \bigg\}
\end{align}

The covering lemma is a powerful tool in classical information theory \cite{CsiszarKorner:82b}. 
\begin{lemma}[Classical Covering Lemma {\cite{CsiszarKorner:82b}\cite[Lemma 3.3]{ElGamalKim:11b}}]
\label{lemm:covering}
Let $X^n\sim \prod_{i=1}^n p_X(x_i)$, $\delta>0$, and let $Z^n(m)$, $m\in [1: 2^{nR}]$, be %
independent random sequences distributed according to $\prod_{i=1}^n p_{Z}(z_i)$. Suppose that the sequence $X^n$ is pairwise independent of the sequences $Z^n(m)$, $m\in [1:2^{nR}]$. Then,
\begin{align}
\prob{ (Z^n(m),X^n)\notin\tset(p_{Z,X}) \,\text{for all $m\in [1: 2^{nR}]$}  }\leq \exp\left( -2^{n(R- I(Z;X)-\eps_n(\delta )} \right)
\end{align}
where $\eps_n(\delta)$ tends to zero as $n\rightarrow\infty$ and $\delta\rightarrow 0$.
\end{lemma}
Let $X^n\sim \prod_{i=1}^n p_X(x_i)$ be an information source sequence,  encoded by an index $m$ at compression rate $R$.
Based on the covering lemma above, as long as the compression rate is higher than $I(Z;X)$,
a set of random codewords, $ Z^n(m)\sim \prod_{i=1}^n p_Z(z_i) $, contains with high probability at least one sequence that is jointly typical with the source sequence.

Though originally stated in the context of lossy source coding, the classical covering lemma is useful in a variety of scenarios
\cite{ElGamalKim:11b}, including communication with CSI
\cite{Pereg:21p}. In our analysis in the sequel, we will have a measurement sequence 
$S^n$ playing the role of the ``source sequence".

\subsection{Quantum Typical Subspaces}
Moving to the quantum method of types, 
suppose that the state of a system is generated from an ensemble $\{ p_X(x), |x\rangle \}_{x\in\Xset}$, hence, the average density operator is
\begin{align}
\rho=\sum_{x\in\Xset} p_X(x) \kb{x} \,.
\end{align}
Consider the  subspace  spanned by the vectors $| x^n \rangle$, for $x^n\in\tset(p_X)$. %
The projector onto the subspace %
is defined as
\begin{align}
\Pi^\delta(\rho)\equiv \sum_{x^n\in\tset(p_X)} \kb{ x^n } \,.
\end{align}
Based on \cite{Schumacher:95p} \cite[Theorem 12.5]{NielsenChuang:02b}, for every $\eps,\delta>0$ and sufficiently large $n$, the $\delta$-typical projector satisfies
\begin{align}
\trace( \Pi^\delta(\rho) \rho^{\otimes n} )\geq& 1-\eps  \label{eq:UnitT} \\
 2^{-n(H(\rho)+c\delta)} \Pi^\delta(\rho) \preceq& \,\Pi^\delta(\rho) \,\rho^{\otimes n}\, \Pi^\delta(\rho) \,
\preceq 2^{-n(H(\rho)-c\delta)}
\label{eq:rhonProjIneq}
\\
\trace( \Pi^\delta(\rho))\leq& 2^{n(H(\rho)+c\delta)} \label{eq:Pidim}
\end{align}
where $c>0$ is a constant.

We will also need the conditional $\delta$-typical subspace. Consider a  state
\begin{align}
\sigma=\sum_{x\in\Yset} p_{X}(x) \rho_B^x 
\end{align}
with 
\begin{align}
 \rho_B^x =\sum_{y\in\Yset} p_{Y|X}(y|x) \kb{\psi^{x,y}} \,.
\end{align}
Given a fixed sequence $x^n\in\Xset^n$, divide the index set $[1:n]$ into the subsets $I_n(a)=\{ i: x_i=a  \}$, $a\in\Xset$,
and define the conditional $\delta$-typical subspace $\mathscr{S}^\delta(\sigma_B|x^n)$ as the span of the vectors
$|\psi^{x^n,y^n}\rangle=\otimes_{i=1}^n |\psi^{x_i,y_i}\rangle$ such that %
\begin{align}
y^{I_n(a)}\in \Aset_\delta^{(|I_n(a)|)}(p_{Y|X=a}) \,,\;\text{for $a\in\Xset$}\,.
\end{align}
The projector onto the conditional $\delta$-typical subspace is defined as
\begin{align}
\Pi^\delta(\sigma_B|x^n)\equiv %
\sum_{|\psi^{x^n,y^n}\rangle\in\mathscr{S}^\delta(\sigma_B|x^n)} \kb{ \psi^{x^n,y^n} }
 \,.
\end{align}
Based on \cite{Schumacher:95p} \cite[Section 15.2.4]{Wilde:17b}, for every $\eps',\delta>0$ and sufficiently large $n$, %
\begin{align}
\trace( \Pi^\delta(\sigma_B|x^n) \rho_{B^n}^{x^ n} )\geq& 1-\eps'  \label{eq:UnitTCond} \\
 2^{-n(H(B|X')_\sigma+c'\delta)} \Pi^\delta(\sigma_B|x^n) \preceq& \,\Pi^\delta(\sigma_B|x^n) \,\rho_{B^n}^{x^ n}\, \Pi^\delta(\sigma_B|x^n) \,
\preceq 2^{-n(H(B|X')_{\sigma}-c'\delta)}
\label{eq:rhonProjIneqCond}
\\
\trace( \Pi^\delta(\sigma_B|x^n))\leq& 2^{n(H(B|X')_\sigma+c'\delta)} \label{eq:PidimCond}
\end{align}
where $c'>0$ is a constant, $\rho_{B^n}^{x^n}=\bigotimes_{i=1}^n \rho_{B_i}^{x_i}$, and the classical random variable $X'$ is distributed according to the type of $x^n$.
Furthermore, if $x^n\in\tset(p_X)$, then %
\begin{align}
\trace( \Pi^\delta(\sigma_B) \rho_{B^n}^{x^n} )\geq& 1-\eps' \,. 
\label{eq:UnitTCondB}
\end{align}
 (see \cite[Property 15.2.7]{Wilde:17b}).

\subsection{Quantum Packing Lemma}
To prove achievability for the HSW Theorem (see Theorem~\ref{theo:ClNoSI}), one may invoke the quantum packing lemma \cite{HsiehDevetakWinter:08p,Wilde:17b}.
Suppose that Alice employs a codebook that consists  of $2^{nR}$ codewords $x^n(m)$, $m\in [1:2^{nR}]$, by which she chooses a quantum state from an ensemble $\{\rho_{x^n} \}_{x^n\in\Xset^n}$. The proof is based on random codebook generation, where the codewords are drawn at random according to an input distribution $p_X(x)$. To recover the transmitted message, Bob may perform the square-root measurement \cite{Holevo:98p,SchumacherWestmoreland:97p} using
a code projector $\Pi$ and codeword projectors $\Pi_{x^n}$, $x^n\in\Xset^n$, which project onto subspaces of the Hilbert space 
$\Hset_{B^n}$. 
The lemma below is a  simplified, less general, version of the quantum packing lemma by Hsieh, Devetak, and Winter \cite{HsiehDevetakWinter:08p}.
\begin{lemma}[Quantum Packing Lemma {\cite[Lemma 2]{HsiehDevetakWinter:08p}}]
\label{lemm:Qpacking}
Let %
\begin{align}
\rho=\sum_{x\in\Xset} p_X(x) \rho_x
\end{align}
where $\{ p_X(x), \rho_x \}_{x\in\Xset}$ is a given ensemble. %
Furthermore, suppose that there is  a code projector $\Pi$ and codeword projectors $\Pi_{x^n}$, $x^n\in\tset(p_X)$, that satisfy for every 
$\alpha>0$ and sufficiently large $n$,
\begin{align}
\trace(\Pi\rho_{x^n})\geq&\, 1-\alpha \\
\trace(\Pi_{x^n}\rho_{x^n})\geq&\, 1-\alpha \\
\trace(\Pi_{x^n})\leq&\, 2^{n d}\\
\Pi \rho^{\otimes n} \Pi \preceq&\, 2^{-n(D-\alpha)} \Pi 
\end{align}
for some $0<d<D$ with $\rho_{x^n}\equiv \bigotimes_{i=1}^n \rho_{x_i}$.
Then, there exist codewords $x^n(m)$, $m\in [1:2^{nR}]$, and  a POVM $\{ \Lambda_m \}_{m\in [1:2^{nR}]}$ such that 
\begin{align}
\label{eq:QpackB}
  \trace\left( \Lambda_m \rho_{x^n(m)} \right)  \geq 1-2^{-n[ D-d-R-\eps_n(\alpha)]}
\end{align}
for all %
$m\in [1:2^{nR}]$, where $\eps_n(\alpha)$ tends to zero as $n\rightarrow\infty$ and $\alpha\rightarrow 0$. 
\end{lemma}
In our analysis, where there is CSI at the encoder, we  apply the packing lemma such that the quantum ensemble encodes both the message $m$ and a compressed representation of the parameter sequence $s^n$.

\section{Proof of Theorem~\ref{theo:MskCl}}
\label{app:MskCl}

Consider a quantum state-dependent channel $\channel_{EA\rightarrow B}$ with CSI at the encoder.

\subsection*{Part 1}
\subsection{Direct Part}
We show that for every $\zeta_0,\eps_0,\delta_0>0$, there exists a $(2^{n(R-\zeta_0)},n,\eps_0,L+\delta_0)$ code for $\channel_{EA\rightarrow B}$, provided that $(R,L)\in \mathcal{R}_{\text{Cl}}(\channel)$. 
To prove achievability, we extend  the classical binning technique and  apply the quantum packing lemma and classical covering lemma. 

The code construction, encoding and decoding procedures are described below.

\subsubsection{Classical Codebook Construction}
Let $\delta>0$, and let $\tR>R$ be chosen later. 
We construct $2^{nR}$ sub-codebooks at random. 
For every message $m\in [1:2^{nR}]$, select $2^{n(\tR-R)}$ independent sequences $x^n(k)$ at random,
each according to $\prod_{i=1}^n p_X(x_i)$. Then, we have the following sub-codebooks,
\begin{align}
\mathscr{B}(m)=\{ x^n(k) \,:\; k\in [ (m-1)2^{n(\tR-R)}+1: m2^{n(\tR-R)}] \}
\end{align}
for $m\in [1:2^{nR}]$.

\subsubsection{Encoding and Decoding}
To send a message $m$,  Alice performs the following.
\begin{enumerate}[(i)]
\item
Measure the CSI systems $E_{0,i}$ using the POVM $\Lambda_{E_0}^s$, for $i\in [1:n]$. Since the CSI systems are in a product state, the measurement outcome is an i.i.d. sequence $\sim q(s)$, where $q(s)=\trace(\Lambda_{E_0}^s\sigma_{E_0})$.

\item
Given a measurement outcome $s^n$, find a sequence $x^n(k)\in \mathscr{B}(m)$ such that $ (s^n, x^n(k))\in \tset(p_{S,X}) $, where
$p_{S,X}(s,u)=q(s)p_{X|S}(u|s) $. If there is none,  select $x^n(k)$ arbitrarily, and if there is more than one such sequence, choose the first among them.

\item
Transmit $\rho^{m}_{A^n}= \bigotimes_{i=1}^n \varphi_A^{x_{i}(k) }$.
\end{enumerate}

Bob receives the output system $B^n$, such that
\begin{align}
\rho_{B^{n}}^m=  \bigotimes_{i=1}^n \rho_B^{x_{i}(k)}
\label{eq:rhoBTnSC}
\end{align}
 and decodes $\hat{k}$  by applying a POVM $\{ \Lambda_{k} \}_{ k \in  [1:2^{n\tR}]}$, which will be specified later. He declares his estimate $\hm$ to be the corresponding sub-codebook index, \ie  $\hm$ such that $x^n(\hat{k})\in \mathscr{B}(\hm)$.

\subsubsection*{Analysis of Probability of Error and Leakage}
First, we show that the probability of decoding error tends to zero as $n\to\infty$. 
By symmetry, we may assume without loss of generality that Alice sends the message $M=1$ using $K$.
Consider the following events,
\begin{align}
\mathscr{E}_1=& \{  (S^n,X^n(k'))\notin \tset(p_{S,X})  \,,\;\text{for all $k'\in \mathscr{B}(1) $} \} \\
\mathscr{E}_2=& \{  \hat{K}\neq K  \}
\end{align}
By the union of events bound, the probability of error is bounded by
\begin{align}
P_{e}^{(n)}(\Tset,\Fset,\Dset) \leq& %
 \prob{ \mathscr{E}_1 }%
+\cprob{ \mathscr{E}_2 }{ \mathscr{E}_1^c } %
\label{eq:PeBnc}
\end{align}
where the conditioning on $M=1$ is omitted for convenience of notation.
By the classical covering lemma, Lemma~\ref{lemm:covering}, the first term tends to zero as $n\rightarrow\infty$ for
\begin{align}
\tR-R> I(X;S)+\eps_1(\delta) \,.
\label{eq:B1nc}
\end{align}
Hence, we choose
\begin{align}
\tR= R+I(X;S)+2\eps_1(\delta) \,.
\label{eq:B1nc2}
\end{align}

To bound the second term, we use the quantum packing lemma. Given $\mathscr{E}_1^c$, we have  $X^n(K)\in\Aset^{\delta_1}(p_X)$, with $\delta_1\triangleq \delta|\Sset| $. Next, observe that
\begin{align}
\Pi^{\delta}(\rho_B)  \rho_{B^n}   \Pi^{\delta}(\rho_B) \preceq& 2^{ -n(H(B)_{\rho}-\eps_2(\delta)) } \Pi^{\delta}(\rho_B)
\\
\trace\left[ \Pi^{\delta}(\rho_B|x^n) \rho_{B^n}^{x^n} \right] \geq& 1-\eps_2(\delta) \\
\trace\left[ \Pi^{\delta}(\rho_B|x^n)  \right] \leq& 2^{ n(H(B|X)_{\rho} +\eps_2(\delta))} \\
\trace\left[ \Pi^{\delta}(\rho_B) \rho_{B^n}^{x^n} \right] \geq& 1-\eps_2(\delta) 
\end{align}
for $x^n\in\Aset^{\delta_1}(p_X)$, by (\ref{eq:rhonProjIneq}), (\ref{eq:UnitTCond}),  (\ref{eq:PidimCond}), and 
(\ref{eq:UnitTCondB}), respectively. Thus, by Lemma~\ref{lemm:Qpacking}, there exists a POVM $D_{k}$ such that
the second error term in (\ref{eq:PeBnc}) is bounded by
$%
\cprob{ \mathscr{E}_2 }{ \mathscr{E}_1^c } \leq 2^{ -n( I(X;B)_\rho -\tR-\eps_3(\delta)) } 
$, %
which tends to zero as $n\rightarrow\infty$, if
\begin{align}
\tR< I(X;B)_\rho -\eps_3(\delta) \,.
\label{eq:B2nc}
\end{align}
Hence, by (\ref{eq:B1nc2}), the probability of decoding error tends to zero, provided that the transmission rate is bounded by
\begin{align}
R< I(X;B)_\rho-I(X;S) -\eps_3(\delta)-2\eps_2(\delta) \,.
\label{eq:B2nc2}
\end{align}

As for the leakage rate, observe that
\begin{align}
I(C^n;B^n)_\rho&\leq I(C^n;X^n(K),B^n)_\rho \nonumber\\
&= I(C^n;X^n(K))_\rho+I(C^n;B^n|X^n(K))_\rho \,.
\label{eq:DirLeak1}
\end{align}
Then,  the first term is bounded by
\begin{align}
I(C^n;X^n(K))_\rho &\leq I(C^n;M,X^n(K))_\rho \nonumber\\
&\stackrel{(a)}{=} I(C^n;X^n(K)|M)_\rho \nonumber\\
&\leq H(X^n(K)|M)_\rho \nonumber\\
&\stackrel{(b)}{\leq} n(\tR-R) \nonumber\\
&\stackrel{(c)}{=} n(I(X;S)+2\eps_1(\delta)) \nonumber\\
&\leq n(I(X;C,S)+2\eps_1(\delta))
\label{eq:DirLeak2}
\end{align}
where $(a)$ holds since $I(C^n;M)_\rho=0$, as there is no correlation between the classical message $M$ and the channel state system $C^n$, $(b)$ follows as $X^n(K)$ belongs to a sub-codebook $\mathscr{B}(M)$ of size $2^{n(\tR-R)}$, and $(c)$ is due to (\ref{eq:B1nc2}).
Moving to the second term in the RHS of (\ref{eq:DirLeak1}),
\begin{align}
I(C^n;B^n|X^n(K))_\rho &\leq I(C^n,S^n;B^n|X^n(K))_\rho
\nonumber\\
&=H(B^n|X^n(K))_\rho-H(B^n|C^n, S^n, X^n(K))_\rho
\label{eq:DirLeak3}
\end{align}
Now, since conditioning does not increase the quantum entropy,
\begin{align}
H(B^n|X^n(K))_\rho\leq \sum_{i=1}^n H(B_i|X_i(K))_\rho=nH(B|X)_\rho \,.
\label{eq:DirLeak4}
\end{align}
Furthermore, given $X^n(K)=x^n$ and $S^n=s^n$, we have a product output state $\rho_{B^n C^n}\equiv \bigotimes_{i=1}^n \channel_{EA\to B}(\sigma_{EC}^{s_i}\otimes \varphi_A^{x_i,s_i})$, where $\sigma^s_{EC}$ denotes  the post-measurement state, \ie $\sigma^s_{EC}\equiv \trace_{E_0}(\Lambda_{E_0}^s\phi_{E_0 E C})/\trace(\Lambda_{E_0}^s\phi_{E_0})$ for $s\in\Sset$. Thus,
\begin{align}
H(B^n|C^n, S^n, X^n(K))_\rho=nH(B|C,S,X)_\rho \,.
\label{eq:DirLeak5}
\end{align}

It follows from (\ref{eq:DirLeak1})-(\ref{eq:DirLeak5}) that
\begin{align}
\frac{1}{n}I(B^n;C^n)&\leq I(X;C,S)+2\eps_1(\delta)+H(B|X)_\rho-H(B|C,S,X)_\rho \nonumber\\
&=I(C,S;X)+I(C,S;B|X)+2\eps_1(\delta)\nonumber\\
&=I(C,S;X,B)+2\eps_1(\delta) \,.
\end{align}
Thereby, the leakage requirement holds if
\begin{align}
I(C,S;X,B)\leq L-2\eps_1(\delta) \,.
\end{align}
To show that rate-leakage pairs in the regularized formula, $\frac{1}{\kappa}\mathcal{R}_{\text{Cl}}(\channel^{\otimes \kappa})$, are achievable as well, one may 
use the coding scheme above over the product channel $\channel^{\otimes \kappa}$, where $\kappa$ is arbitrarily large.
This completes the proof of the direct part.

\subsection{Converse Part}
The proof of the regularized converse part is a straightforward extension of standard considerations. For completeness, we give the details below. 
Suppose that Alice and Bob are trying to distribute randomness. An upper bound on the rate at which Alice can distribute randomness to Bob also serves as an upper bound on the classical communication rate. Then, suppose that Alice prepares a maximally correlated state
\begin{align}
\pi_{MM'} \equiv \frac{1}{2^{nR}}\sum_{m=1}^{2^{nR}} \kb{ m }_M \otimes \kb{ m }_{M'} .
\end{align}
locally, where $M$ and $M'$ are classical message registers. Denote the joint state at the beginning by
\begin{align}
\psi_{MM'   E_0^n E^n C^n}= \pi_{MM'}\otimes \phi_{ E_0 E C}^{\otimes n}
\end{align}
where %
$E^n$ are the channel state systems, $E_0^n$ are the CSI systems that are available to Alice, and $C^n$ are the systems that are masked from Bob (see Figure~\ref{fig:mskCl1}).

 Alice performs a measurement $\Tset_{E_0^n\to V}$ on the CSI systems $E_0^n$, and obtains a measurement outcome $V$.
Denote the average post-measurement state by
\begin{align}
 \rho_{MM'   V E^n C^n}\equiv \Tset_{E_0^n \rightarrow V}( \psi_{MM' E_0^n   E^n  C^n}) \,.
\end{align}
 Then, she applies an encoding map $\Fset_{M' V\to A^n V}$ to the classical system $M'$ and the measurement outcome $V$ (since $V$ is classical, it can be copied.) %
The resulting  state is 
\begin{align}
\rho_{M  A^n V  E^n C^n}\equiv \Fset_{M' V \rightarrow A^n V}( \rho_{MM' V   E^n  C^n} %
) .
\end{align}
As the input systems $A^n$ are sent through the channel, the output state is
\begin{align}
\rho_{M B^n   C^n V}\equiv \channel^{\otimes n}_{E A\rightarrow B} (\rho_{  M E^n A^n  C^n V}) .
\end{align}
Bob receives $B^n$ and performs a decoding channel $\Dset_{B^n \rightarrow \hM}$, producing 
\begin{align}
\rho_{  M \hM  C^n V}\equiv \Dset_{B^n \rightarrow \hM}(\rho_{M B^n  C^n V}) .
\label{eq:DecConv1}
\end{align}

Consider a sequence of codes $(\Tset_n,\Fset_n,\Dset_n)$ %
such that
\begin{align}
\frac{1}{2} \norm{ \rho_{M\hM} -\pi_{MM'} }_1 &\leq \eps_n \label{eq:randDconv} \\
\frac{1}{n} I(C^n V;B^n)_\rho &\leq L+\delta_n \label{eq:randDconvLeak}
\end{align}
where %
$\eps_n,\delta_n$ tend to zero as $n\rightarrow\infty$.
Based on the Alicki-Fannes-Winter inequality \cite{%
Winter:16p} \cite[Theorem 11.10.3]{Wilde:17b}, (\ref{eq:randDconv}) implies 
\begin{align}
|H(M|\hM)_\rho - H(M|M')_{\pi} |\leq n\eps_n'
\label{eq:AFW}
\end{align}
where $\eps_n'\to 0$  as $n\rightarrow\infty$.
Since $H(\pi_{M M'})=H(\pi_{M})=H(\pi_{ M'})=nR$, we have $I(M;\hM)_{\pi}=nR$.
Then, as   $H(\rho_{M})=H(\pi_{ M})=nR$, we also have
$I(M;M')_{\pi} - I(M;\hM)_{\rho}= H(M|\hM)_\rho - H(M|M')_{\pi}$. Thus,  (\ref{eq:AFW}) implies
\begin{align}
nR&=I(M;\hM)_{\pi} \nonumber\\
&\leq I(M;\hM)_{\rho}+n\eps_n' \nonumber\\
&\leq I(M;B^n)_{\rho}+n\eps_n '
\label{eq:ConvIneq1}
\end{align}
where the last line follows from (\ref{eq:DecConv1}) and the quantum data processing inequality \cite[Theorem 11.5]{NielsenChuang:02b}.
Since the message has no correlation with the channel state system $E_0^n$, we can also write this as
\begin{align}
nR&\leq I(M;B^n)_{\rho}-I(M;V)_{\rho}+n\eps_n'
\nonumber\\
&=I(X^n;B^n)_{\rho}-I(M;S^n)_{\rho}+n\eps_n'
\label{eq:ConvIneq1b}
\end{align}
as we define $X^n=f(M)$ and $S^n=g(V)$, where $f$ and $g$ are arbitrary one-to-one maps.
This concludes the converse proof for part 1.

\subsection*{Part 2}
Now, we consider the %
special case of a measurement channel $\Mset_{EA\rightarrow Y}$, where the CSI system and the channel output are classical, \ie $E_0\equiv S\sim q(s)$ and $B\equiv Y$. The direct part follows from 
part 1. %
To prove the converse part, we extend the methods of Merhav and Shamai  \cite{MerhavShamai:07p}. 

By the classical chain rule,
\begin{align}
I( M;Y^n)&= \sum_{i=1}^n I(M;Y_i|Y^{i-1})
\nonumber\\
&= \sum_{i=1}^n I( M Y^{i-1} S_{i+1}^n;Y_i) %
-\sum_{i=1}^n I(Y_i;S_{i+1}^n| M Y^{i-1}) \nonumber\\
&= \sum_{i=1}^n I( M Y^{i-1} S_{i+1}^n;Y_i) %
-\sum_{i=1}^n I(Y^{i-1}; S_i| M  S_{i+1}^n )
\label{eq:ConvIneq3}
\end{align}
where the last line follows from the %
Csisz\'ar sum identity \cite[Section 2.3]{ElGamalKim:11b}.  %
Since  $ S_i$ and $(M, S_{i+1}^n)$ are statistically independent,  we have $I(Y^{i-1}; S_i| M  S_{i+1}^n )=
I(  M  S_{i+1}^n Y^{i-1}; S_i)$. 
Therefore, defining  
\begin{align}
X_i=( M,Y^{i-1}, S_{i+1}^n)
\label{eq:EAconvAi}
\end{align}
we obtain 
\begin{align}
I(  M;Y^n)_{\rho}&\leq \sum_{i=1}^n I(X_i;Y_i)_\rho -\sum_{i=1}^n I(X_i; S_i)_\rho .
\label{eq:ConvIneq4}
\end{align}

Let  $J$ be a classical random variable with a uniform distribution over
$\{ 1,\ldots, n \}$, in a product state with the previous quantum systems, \ie $ C^n$, $E^n$, $E_0^n$,  $M$, $M'$, $A^n$, and
$Y^n$. 
Then, by (\ref{eq:ConvIneq1}) and (\ref{eq:ConvIneq4}), 
\begin{align}
R-\eps_n'%
&\leq 
\frac{1}{n}\sum_{i=1}^n [ I(X_i;Y_i)_\rho-I(X_i;S_i)_\rho ] 
\nonumber\\
&= I(X_J;Y_J|J)-I(X_J;S_J|J)  \nonumber\\
&= I(X_J,J;Y_J)-I(J;Y_J)%
-I(X_J,J;S_J)_\rho+I(J;S_J)_\rho \nonumber\\
&\leq I(X_J,J;Y_J)_\rho-I(X_J,J;S_J)_\rho+I(J;S_J)_\rho \nonumber\\
&= I(X_J,J;Y_J)_\rho-I(X_J,J;S_J)_\rho
\end{align}
with %
$%
\rho_{J X_J E_J C_J  A_J}=\frac{1}{n} \sum_{i=1}^n \kb{i} \otimes \rho_{X_i E_i C_i  A_i} %
$ and $%
\rho_{J X_J C_J Y_J }=\Mset_{EA\rightarrow Y}(\rho_{J X_J  C_J E_J A_J}) %
$, %
where the last equality holds since the sequence $ S^n$ is  i.i.d. Thus, defining
\begin{align}
X\equiv (X_J,J) \,,\; S\equiv S_J \,,\; E\equiv E_J \,,\;  C\equiv C_J \,,\; A\equiv A_J 
\label{eq:ConvDsin}
\end{align}
 and $Y$ such that 
$%
\rho_{YC}=\Mset_{EA\rightarrow Y}(\rho_{EA C}) %
$, we obtain the desired bound on the coding rate,
\begin{align}
R-\eps_n'\leq I(X;Y)-I(X;S) \,.
\end{align}

As for the leakage rate, by (\ref{eq:randDconvLeak}),
\begin{align}
n(L+\delta_n) &\geq   I( C^n S^n;Y^n )_\rho 
\nonumber\\
&=  I(C^n S^n;Y^n M)_\rho -I(C^n S^n; M |Y^n )_\rho 
\nonumber\\
&= I(C^n S^n;Y^n    M)_\rho -H( M |Y^n )_\rho %
+ H( M |C^n S^n Y^n )_\rho .
\label{eq:ineqL1}
\end{align}
For  a classical-quantum state $\rho_{XA}=\sum_{x\in\Xset}p_X(x)\kb{x}\otimes \rho_A^x$,
 the conditional entropy of  is always  nonnegative, as %
$H(X|A)_\rho\geq H(X|A,X)=0$, since conditioning cannot increase quantum entropy \cite[Theorem 11.15]{NielsenChuang:02b}. As the message $M$ is  classical, the last term in the RHS of (\ref{eq:ineqL1}) is nonnegative, i.e. 
\begin{align}
H( M |C^n, Y^n)_\rho\geq 0 .
\label{eq:ineqL1a}
\end{align}
Furthermore,  by (\ref{eq:ConvIneq1}),  the second term satisfies
\begin{align}
H( M |Y^n)_\rho = H(M)_{\pi} - I( M ;Y^n )_\rho\leq n\eps_n' .
\label{eq:ineqL1b}
\end{align}
Thus, by (\ref{eq:ineqL1})-(\ref{eq:ineqL1b}),
 \begin{align}
n(L+\eps_n'+\delta_n) &\geq    I(C^n S^n;Y^n M)_\rho 
\nonumber\\
&= \sum_{i=1}^n I(C_i S_i;Y^n  M|C_{i+1}^n S_{i+1}^n)_\rho 
\nonumber\\
&\geq \sum_{i=1}^n I(C_i S_i;Y_i Y^{i-1}  M|C_{i+1}^n S_{i+1}^n)_\rho .
\end{align}
Then, since  $(C_i,S_i)$ and $(C_{i+1}^n,S_{i+1}^n)$ are in a product state, we have $I(C_i S_i;C_{i+1}^n S_{i+1}^n)_\rho=0$. Hence,
\begin{align}
L+\eps_n'+\delta_n %
&\geq     \frac{1}{n}\sum_{i=1}^n I(C_i S_i;Y_i Y^{i-1}   M C_{i+1}^n S_{i+1}^n)_\rho \nonumber\\
&\geq     \frac{1}{n}\sum_{i=1}^n I(C_i S_i;Y_i Y^{i-1}   M  S_{i+1}^n)_\rho \nonumber\\
&=     \frac{1}{n}\sum_{i=1}^n I(C_i S_i;X_i Y_i)_\rho\nonumber\\
&=      I(C_J S_J; X_J Y_J|J)_\rho \nonumber\\
&=      I(C_J S_J;X_J J Y_J)_\rho\nonumber\\
&=I(C S;X Y)_\rho
\end{align} 
where the first equality follows from our definition of $X_i$ in (\ref{eq:EAconvAi}), the second holds since  $J$ is a the classical variable with a uniform distribution over $\{1,\ldots,n\}$, the third because $I(C_J S_J;J)_\rho=H(C_J S_J)_\rho-H(C_J S_J|J)_\rho=H(C S)_\phi-H(CS)_\phi=0$, and the last equality follows from the definition of $C$, $S$, $X$, and $Y$ in (\ref{eq:ConvDsin}).
This completes the proof of Theorem~\ref{theo:MskCl}.
\qed %

\bibliography{references3}{}

\end{document}